\documentclass[12pt,preprint]{aastex}

\begin{document}

\title{A tidal disruption model for the gamma-ray burst of GRB 060614}

\author{Y. Lu\altaffilmark{1}, Y.F. Huang\altaffilmark{2}, and S.N. Zhang\altaffilmark{3}}
\altaffiltext{1}{National Astronomical Observatories, Chinese
Academy Of Sciences, Beijing 100012, China; ly@bao.ac.cn}
\altaffiltext{2}{Department of Astronomy, Nanjing University,
Nanjing 210093, China}
 \altaffiltext{3}{Physics Department and
Center for Astrophysics, Tsinghua University, Beijing 100084, China}
% \altaffiltext{4}{Department of Physics, The
% University of Hong Kong, Pokfulam Road, Hong Kong, China}

\begin{abstract}
The combination of a long duration and the absence of any
accompanying supernova clearly shows that GRB 060614 can not be
grouped into the two conventional classes of gamma-ray bursts, i.e.
the long/soft bursts deemed to be collapsars and the short/hard
bursts deemed to be merging binary compact stars. A new progenitor
model is required for this anomalous gamma-ray burst. We propose
that GRB 060614 might be produced through the tidal disruption of a
star by an intermediate mass black hole. In this scenario, the long
duration and the lack of any associated supernova are naturally
expected. The theoretical energy output is also consistent with
observations. The observed 9-s periodicity in the $\gamma$-ray light
curve of GRB 060614 can also be satisfactorily explained.

\end{abstract}


\section{Introduction}

Gamma-ray bursts (GRBs) are generally grouped into two classes,
namely long/soft bursts and short/hard bursts, separated at
$T_{90}\sim 2\,s$ \citep{Kou93}. Long GRBs are believed to originate
from the collapse of massive stars \citep{Woo93}. They are usually
associated with supernovae \citep{Gal98, Pia06}, and occur in star
forming regions \citep{Fru06}. Short GRBs are widely believed to be
connected with the merge of binary compact stars, and their host
galaxies are thought to have  a lower star formation rate
\citep{Bli84, Eic89, Vil05, Blo06}.  They should not be associated
with any supernovae. Recently, by analogy with the SN
classification, \cite{Zha07} suggested a new GRB terminology scheme:
Type I and Type II bursts. Type I bursts correspond to the star
merging group, and Type II bursts correspond to the collapse of
massive stars.

GRB 060614 is a very special event that has strongly challenged our
classical classification of GRBs (Gehrels et al. 2006).
It has a long duration of $\sim$ 102
s (Gehrels et al. 2006), which seems to suggest  it as a long/soft
GRB. However, no supernova signature has been observed from this
event, although it is very near to us (Gal-Yam et al. 2006; Fynbo
et al. 2006), which is strongly contradicted with a collapsar origin.
Obviously, this GRB cannot be simply classified as
any of the conventional class, either long/soft GRBs or short/hard
GRBs. A novel mechanism is needed for this special burst (Gal-Yam
et al. 2006; Zhang et al. 2007; Mangano et al. 2007; Jakobsson \&
Fynbo 2007; Amati et al. 2007).

Although the formation mechanism of intermediate mass black holes
(IMBHs) is still not clear so far \citep{Heg02, Por04, Mil02,
Geb05}, some evidence has been accumulated for their existence
in galaxies of all types, including dwarf
galaxies, especially in young and global clusters \citep{Map07}. We
propose here that the tidal disruption of a star by an IMBH could be
an ideal way to power a nearby long-duration GRB without any
associated supernova, such as GRB 060614. Our motivation is that
the debris of the tidally disrupted star is likely to form an
accretion disk surrounding the black hole. As a consequence, jets
would be formed in this process \citep{Lu06}, which are similar to
the jets triggered by collapsars and binary mergers. We believe that
the debris accreted onto the IMBH is a conceivable energy source for
GRBs.

The structure of this paper is as follows.
We gather up the observed facts of GRB 060614 in Section 2. Our novel
model for GRB 060614 is described in detail in Section 3. Section 4
is the conclusion and a brief discussion.

\section{Main features of GRB 060614}

GRB 060614 is a very peculiar nearby burst, which was detected by
the Burst Alert Telescope (BAT) onboard the Swift satellite on 2006
June 14 at 12:43:48 UT \citep{Geh06}. The general properties of GRB
060614 can be summarized as follows. (1) It is a long GRB that
lasted for $\sim$ 102 s \citep{Geh06}; (2) An interesting
substructure has been noted in the prompt $\gamma$-ray light curve:
BAT records reveal a first short episode of emission lasting for
$\sim 4$ s followed by an extended and somewhat softer episode
lasting for $\sim 100$ s \citep{Geh06, Fyn06}. A close examination
shows that this short episode is actually composed of $\sim 5$
mini-pulses. It is also noted that a 9-s periodicity exists between
7 and 50 s in the $\gamma$-ray light curve, although it is not
statistically significant (Gehrels et al. 2006). (3) It has a long
duration, but it lies in the region of short/hard GRBs on the
temporal lag-peak luminosity plane (Gehrels et al. 2006); (4) It is
not associated with any supernova, although it is at a relatively
low redshift of $z=0.125$ \citep{Gal06, Fyn06}; (5) Its host is a
faint dwarf galaxy, with a low star formation rate of 0.0084 ---
0.014 M$_\odot$/yr \citep{Jak07, Gal06, Fyn06}; (6) The GRB is
offset from the nucleus of the host galaxy by 0.5'' \citep{Geh06,
Gal06}; (7) The gamma-ray fluence is $\sim 2\times 10^{-5}$
ergs/cm$^{2}$, which corresponds to an isotropic gamma-ray energy
release of $1.08\times 10^{51}$\,ergs in 1 keV --- 10 MeV range in
the GRB restframe \citep{Geh06}. Note that if the radiation
efficiency is 0.1, then the isotropic kinetic energy is $\sim
1\times 10^{52}$ ergs.

The long duration of GRB 060614 suggests it as a long/soft GRB.
However, the absence of any associated supernova signature of such a
nearby event strongly contradicts a collapsar origin. The above
observational facts indicate that GRB 060614 does not fit into any
of the two conventional classes of GRBs, i.e. long/soft and
short/hard GRBs. A completely new type of engine other than the
collapsars or binary compact star mergers is need for this event.
Below, we will show that the tidal disruption of a star by an IMBH
can hopefully explain all the basic features of GRB 060614.

\section{Tidal disruption model for GRB 060614}

There are three kinds of black holes: stellar mass black holes,
intermediate mass black holes (IMBHs), and super-massive black
holes. Studies have been abundant for stellar mass black holes and
super-massive black holes. Stellar mass black holes, with masses
ranging from 3 to 20 M$_\odot$ \citep{Oro03}, are thought to be the
relics of massive stars. Super-massive black holes, with a mass of
$M_{\rm bh} \sim 10^6$ --- $10^9$ M$_\odot$, are believed to reside
at the center of many galaxies. Strong observational evidence for
the existence of stellar mass black holes and super-massive black
holes has been accumulated today. For example, tidal disruption of a
star by a super-massive black hole is expected to produce a luminous
flare of electromagnetic radiation in the UV to X-ray bands at the
center of many galaxies \citep{Ree88, kom99, Ulm99, kom04, Mil06}.
This process is also suggested as a possible source for the observed
non-thermal TeV gamma-ray emission from the center region of our
Galaxy \citep{Lu06}. Interestingly, as one potential model for long
GRBs \citep{Tri02}, the tidal disruption of a star by a
super-massive black hole has already been discussed in the earlier
works of \cite{Car92} and \cite{Che01}. Furthermore, a possibility
of gamma-ray flares linked to shock waves in tidally compressed
stars by massive black holes has been predicted in the more recent
work of \cite{Bra07}.

The masses of IMBHs are in the range of $M_{\rm bh}
\sim $ 20 --- $10^5$ M$_\odot$ \citep{Map06}.  Their existence has
also been inferred from recent observations. For example, the
capture of stars by IMBHs with masses of 350 --- 1200 M$_\odot$ are
suggested as the mechanism for some ultraluminous X-ray sources
\citep{Bau06}. In this section, we investigate the possibility of
producing a GRB 060614-like GRB through the tidal disruption of a star
by an IMBH. The main difference between this GRB model and the
former one proposed by \cite{Che01} is the mass of black holes and
the location in the host galaxies. Note that IMBHs can be offset
from the center region of their host galaxy.

\subsection{General picture}

Despite the great difference between the black hole masses, tidal
disruption of a star by an IMBH should be more or less similar to
the disruption of a star by a super-massive black hole. A transient
accretion disk may be formed in the process. The accretion rate
should be very high (beyond the Eddington rate) at first, but
should decrease steadily with time. As long as the accretion rate
is high enough (nearly the Eddington rate), the inner region of
the disk should be dominated by radiation pressure. The disk
can then anchor and amplify the seed magnetic field ($B_{\rm eq}$)
to a strong ordered poloidal field $(B_{\rm p})$, which in turn
threads the black hole with a mass-flow ring in the inner region
of the disk. A large amount of rotational energy of the black
hole can be extracted via the Blandford-Znajek (BZ) process, creating
two counter-moving jets along the rotation axis \citep{Bla77}.
The powerful jets are ideal energy reservoirs for GRBs.

A jet produced in such a way should be highly variable. The
exponential increase of $B_{\rm eq}$ to $B_{\rm p}$ is linked to an
instability of the disk in the case of a high accretion rate. When
$B_{\rm p}$ is so strong that it dominates over the material pressure,
all debris material within the spherization radius ($R_{\rm sp}$)
\citep{Sha73} is likely to be broken into blobs and fall
successively into the black hole at the marginally stable radius
($R_{\rm ms}$). The falling of each blobs into the black hole should
lead to a clump-like structure in the jet, which may correspond to a
pulse (a mini-burst) in the light curve of the GRB \citep{Che01}.
The width of a mini-burst $(t_{\rm pulse})$ is determined by the
free-fall timescale ($t_{\rm ff}$). The total duration of the whole GRB
($t_{\rm duration}$) is determined by the time needed for all the debris material
in the range of $R_{\rm ms}<R<R_{\rm sp}$ to fall into the black hole at
$R_{\rm ms}$, which in fact is the instability timescale of the disk
($t_{\rm in}$).

The instability should happen when the accretion rate is near the
Eddington rate. Note that when all the original material in the
range of $R_{\rm ms}<R<R_{\rm sp}$ is accreted by the black hole, the
region cannot be replenished again, due to the combination of
the relatively long accretion timescale and the rapidly decreasing
accretion rate. The central engine is then actually quenched, although
consequent energy injection is still possible due to continuous
accretion at a much lower level.

\subsection{Tidal disruption and the formation of a transient accretion disk}

When a star with a mass of $M_*$ and radius of $R_*$ passes by
a massive black hole with a mass of $M_{\rm bh}$, it would be captured
and eventually tidally disrupted at an average tidal radius of
\begin{eqnarray}
R_{\rm T} &\simeq& 3.25\times 10^{12}r_* m_*^{-1/3} M_5^{1/3}\, {\rm cm}, \,\,
\end{eqnarray}
where $m_*=M_*/M_\odot$, $r_*=R_*/R_\odot$, $R_\odot$ and $M_\odot$
are the solar radius and mass, respectively. For convenience, we
introduce the following dimensionless quantities throughout this
article:
$$M_5=\frac{M_{\rm bh}}{10^5M_\odot}, \dot{m}=\frac{\dot{M}}{\eta \dot{M}_{\rm Edd}},
\hat{r}=\frac{R}{R_{\rm ms}},$$ where $R_{\rm ms}=3R_{\rm s}$ and $R_{\rm s}=3\times
10^{10}M_5$ cm is the Schwarzschild radius, $\dot{M}_{\rm Edd}=3\times
10^{-2}\eta_{0.1}^{-1}M_5\,M_\odot / {\rm yr} $ is the Eddington
accretion rate, $\eta$ is the energy conversion factor and
$\eta_{0.1}=\eta/0.1$.

The strength of a tidal disruption generally depends on the black
hole mass and the penetration factor $\beta$ defined as $\beta\equiv
R_{\rm T}/R_{\rm p}$, where $R_{\rm p}$ is the pericenter of the
star's orbit. The maximum value of $\beta$ is 12 and 56 for $M_{\rm
bh} = 10^6$ and $10^5$ M$_\odot$ Schwarzschild black hole
\citep{Kob04}, respectively. When $\beta\gg 1$, the star within
$R_{\rm T}$ can undergo compression to a highly flattened pancake
configuration \citep{Car83,Lum86,Lum89}. Once the star is totally
disrupted in such a way, the matter with higher angular momentum
will rapidly lag behind the matter with lower angular momentum,
producing a long and thin spiral \citep{Gom05}, and resulting in a
continuous debris accreted onto the black hole. Although it is
difficult to predict what fraction of the initially bound debris
will be accreted, the recent numerical simulations indicate that
approximately $25\%$ --- $50\%$ of the initial stellar mass may
remain bound \citep{Aya00}. At first the accretion rate may be much
higher than the Eddington rate and no accretion disk with a
radiation-dominated inner region can be formed. We will not discuss
this phase since no powerful jet can be launched during this phase.
When the accretion rate decreases to about the Eddingtion rate, a
disk of our interest may be formed. The radius of the disk is
typically comparable to the tidal radius $R_{\rm T}$, and the
evolution of the accretion rate is given by \citep{Ree88}
\begin{eqnarray}
\dot{m}=2.64\times 10^3(\frac{t}{t_{\rm D}})^{-5/3}\,\,,
\end{eqnarray}
where $t_{\rm D}$ is the dynamic or orbit timescale, which is given by
(Sanders \& Van Oosterom 1984)
\begin{eqnarray}
t_{\rm D}=(\frac{GM_{\rm bh}}{R_{\rm T}^3})^{-1/2}=1.6\times
10^{3}r_*^{3/2}m_*^{-1/2}\, {\rm s} \,.
\end{eqnarray}

Assuming that the evolution of the debris disk follows the
behavior of a standard disk, then the infall of matter within a
narrow sector (a mass ring) is a special kind of disk accretion.
The matter moves in a slowly twisting spiral in the plane
perpendicular to the direction of the angular momentum of the mass
flow. The time in which a large fraction of the matter in the disk
can be accreted is \citep{Ulm99}
\begin{eqnarray}
t_{\rm acc}\approx9.47\times 10^7\alpha^{-1}
h_{-2}^{-2}\beta^{-3/2}r_*^{3/2}m_*^{-1/2}\,\,s\,,
\end{eqnarray}
where $\alpha$ is the viscous paramater, ranging as
$0\leq\alpha\leq 1$, $h= 10^{-2} h_{-2} =  H / R$ is the ratio of disk height
($H$) to radius and is approximately one for a thick disk. For
a thin disk, we adopt $h=10^{-2}$ .
In this case, the accretion time is very long. It indicates that
when a region of the disk is suddenly cleaned up due to some
mechanism, it would essentially be impossible to replenish this region
again, due to the relatively long accretion timescale ($t_{\rm acc}$) and the
rapidly decreasing accretion rate.

There are three distinct regions in a standard thin disk, depending
on the sources of opacity and pressure \citep{Sha73}: the inner
region, the middle region and the outer region. We concentrate on the
inner region, which is dominated by radiation pressure
and electron scattering. Note that this region exists only when
the accretion rate is high, i.e. $\dot{m}>\dot{m}_{\rm c}$, where
$\dot{m}_{\rm c}=7.97\times 10^{-3}(\alpha M_5)^{-1/8}$ is a critical
accretion rate \citep{Sha73}. In this case, the disk has a
constant thickness along the radius for $\hat{r}\gg 1$,
and the thickness depends only on the accretion rate. The
maximal value of $h$ is reached at $R_{\rm sp}$ as $h_{\rm max}=\dot{m}$,
where $R_{\rm sp}$ is the radius of spherization
\citep{Sha73}
\begin{eqnarray}
R_{\rm sp}= 2.25\times 10^{11}\dot{m}M_5\,\, {\rm cm} \,.\nonumber
\end{eqnarray}

Within the radius of $R_{\rm sp}$, the disk may suffer an instability,
e.g., due to mass accumulation resulted from the increase of the
viscosity and the local accretion rate. A thermal ionization
instability is eventually developed. Consequently, magnetic
field can be amplified in the process and jets can be launched
via the BZ mechanism.

%%%%%%%%%%%%%%%%%%%%%%%%%55555

\subsection{Energy extraction}

To estimate the powerful energy extracted via BZ process, we need to
calculate the ordered poloidal field. Irrespective of the detailed
field structure, the original magnetic field in the disk can be
estimated by the assumption of energy equipartition: $
B^2_{\rm eq}/(8\pi) \equiv p_{\rm d,max}$, where
$p_{\rm d,max}$ is the maximum pressure of the
disk. For a high accretion rate, the disk is dominated by the
radiation pressure and the maximum pressure in the inner region can
be described following the work of \cite{Sha73},
\begin{eqnarray}
p_{\rm d,max}=9.89\times 10^{9}(\alpha M_5)^{-1}\hat{r}^{-3/2}\, \,{\rm
dyn}\,{\rm cm}^{-2}&,  \dot{m}>\dot{m}_{\rm c}\, .
\end{eqnarray}
Eq.(5) shows that the maximum pressure is a function of
$\hat{r}^{-3/2}$. Since $B_{\rm eq}\propto p_{\rm d,max}^{1/2}$, we have
$B_{\rm eq}\propto \hat{r}^{-3/4}$. $B_{\rm eq}$ acts as a
seed magnetic field. So the seed field is a function of radius.
It increases quickly with the decrease of the radius.

The instability happens in the region of $R_{\rm ms}<R<R_{\rm sp}$
\citep{Sha73, Che01, Lu06}. We refer this region as the instability
region, and designate the corresponding radius as $R_{\rm in}$.
Accordingly the dimensionless radius is
$\hat{r}_{\rm ms}<\hat{r}_{\rm in}<\hat{r}_{\rm sp}$, where
$\hat{r}_{\rm in}=R_{\rm in}/R_{\rm ms}$, and
$\hat{r}_{\rm sp}=R_{\rm sp}/R_{\rm ms}=2.52\dot{m}$. In this region, the debris
matter is likely to be broken into many blobs, and the seed field
threading the block and the disk may be wrapped up tightly, becoming
highly sheared and predominantly azimuthal in orientation. This
leads to a bursting growth of $B_{\rm eq}$ to a strong poloidal magnetic
fields $B_{\rm p}$ \citep{Che01}.
The final amplified strength, $B_{\rm p} (\propto B_{\rm eq})$, and the
corresponding growth timescale ($\triangle t_{\rm p}$) are, respectively
\citep{Che01},
\begin{eqnarray}
B_{\rm p}^2 &\simeq& 5.67\times 10^{-2}n \, {\rm G}^2 , \nonumber \\
\triangle t_{\rm p} &\simeq& 4.5\times
10^{-4}n^{1/8}\alpha^{1/8}M_5^{9/8}\hat{r}_{\rm in}^{3/8}\,\, {\rm s},
\end{eqnarray}
where $n$ (in units of ${\rm cm}^{-3}$) is the number density in
the inner region. It
is given by \citep{Sha73}
\begin{eqnarray}
n=2.79\times 10^{21}(\alpha M_5)^{-1}\hat{r}_{\rm in}^{-9/8}\,\,
& ( \dot{m}>\dot{m}_{\rm c})\,.
\end{eqnarray}
Substituting Eq.(7) into Eq.(6), we have
\begin{eqnarray}
&& B_{\rm p}^2=1.58\times 10^{20}(\alpha
M_5)^{-1}\hat{r}_{\rm in}^{-9/8}\,\, {\rm G}^2, \nonumber\\
&& \triangle t_{\rm p}\simeq 2.15\times
10^{-1}M_5\hat{r}_{\rm in}^{15/64}\,\, {\rm s}.
\end{eqnarray}

In the region of $R_{\rm ms} < R < R_{\rm sp}$, the debris matter is
broken into many blobs. The characteristic dimension for each blob
is limited by the geometrical thickness of the accretion flow, and
thus can be estimated as $\sim H$  \citep{Abr85}. Owing to general
relativistic effects, near the black hole, the relative value of $H$
can be estimated by \citep{Abr85}
\begin{eqnarray}
\frac{H}{R_{\rm in}}\sim 10^{-2}\zeta^{-1/2}\chi\,\, , \nonumber
\end{eqnarray}
where $\chi$ is a parameter depending on the mass of the black hole
and the accretion rate, and $\zeta$ is the ratio of the gas pressure
to the total pressure of the disk. For massive black holes,
$\chi\sim 0.1$, and for a thin disk the value of $\zeta$ ranges as
$10^{-4} \leq \zeta \leq 1$. Here, we adopt $\zeta \sim 10^{-3}$ for the
disk dominated by radiation pressure. Since the instability region
is within $R_{\rm sp}$, the total number of the blobs
appearing in the instability region can be estimated as
\begin{eqnarray}
N_{\rm tot} \sim \frac{R_{\rm in}}{H}\sim
30\zeta^{1/2}_{-3}\chi_{0.1}^{-1}\,,
\end{eqnarray}
where $\chi_{0.1}=\chi/0.1$, and $\zeta_{-3}=\zeta/10^{-3}$. These
blobs can be maintained on a diffusion timescale of $\sim H/r_{\rm
L} c$, where $r_{\rm L}$ is the Larmor radius \citep{Che01}.

When a blob reaches $\hat{r}_{\rm ms}$ and be dragged into the black
hole, a huge amount of energy will be extracted via the BZ process
\citep{Bla77}, giving birth to a mini-burst. This process may repeat
many times till all blobs within $\hat{r}_{\rm sp}$ are removed
(c.f.~Eq.(9)). As a result, many mini-bursts should be produced. We
propose that these mini-bursts add together to form a GRB, and each
mini-burst corresponds to a mini-pulse in the GRB light curve.

The timescale of each mini-pulse is determined by $t_{\rm pulse} \sim
max(\triangle t_{\rm p} \,, t_{\rm ff})$ \citep{Che01}, where $t_{\rm ff}$ is
the free fall time. Assuming that the black hole is accreting at the
Eddington rate ($\dot{m}=1$), we derive the characteristic duration
of the mini-pulse as,
\begin{eqnarray}
t_{\rm pulse}=t_{\rm ff}=3R_{\rm s}/c\simeq 3M_5\, {\rm s} \,,
\end{eqnarray}
which is mainly determined by the black hole mass.

In our model, the total duration of a GRB is determined by the timescale
of the thermal instability, i.e.,
\begin{eqnarray}
t_{\rm duration}=\frac{R_{\rm sp}}{v_{\rm R}}
= 50\alpha^{-1}\hat{r}_{\rm ms}^{3/2}M_5\, {\rm s} \,.
\end{eqnarray}
where $v_{\rm R}$ is the radial velocity of the matter, and $v_{\rm
R}=\alpha \sqrt{GM_{\rm bh}/R_{\rm sp}}$ \citep{Sha73}. In this main
burst phase, the majority of matter within the range of $1 < \hat{r}
< \hat{r}_{\rm sp}$ will fall into the black hole. It will take a
time that equals to the accretion timescale ($t_{\rm acc}$) to
replenish this mass. However, the accretion timescale is very long,
$t_{\rm acc} \sim 3.0 \times 10^6 \alpha^{-1} \, {\rm s}$ by
assuming $\beta=1$ (c.f. Eq.(4)). Within this time, the accretion
rate will decrease to a value (c.f. Eq.(2)) that is much lower than
the critical accretion rate of $\dot{m}_{\rm c}$. Consequently, the
radiation pressure of the disk is too low to produce enough magnetic
field to trigger any additional mini-bursts again \citep{Che01}.

Assuming that the magnetic field component normal to the black hole
horizon equals to the amplified magnetic field ($B_{\rm p})$, we can
further estimate the power and energy extracted via the BZ process
in a single mini-burst \citep{Che01}, respectively
\begin{eqnarray}
&& P_{\rm pulse}=1.7\times 10^{48}A^2f(A)M_5^{2}(\frac{B_{\rm p}}{10^9 {\rm
G}})^{2}\hat{r}_{\rm in}^{-9/8}\,\,\,
{\rm ergs} / {\rm s},\nonumber\\
&& E_{\rm pulse}\simeq P_{\rm pulse}\triangle t_{\rm p} \simeq 2.69\times
10^{50}A^2f(A)\alpha^{-1}M_5\hat{r}_{\rm in}^{-57/64}\, {\rm ergs} \,  ,
\end{eqnarray}
where $A$ is the dimensionless angular momentum of the black hole,
and $f(A)=2/3$ for $A\rightarrow 0$ and $f(A)=\pi-2$ for
A$\rightarrow$ 1 \citep{Lee00}. It has been known that some stellar
mass black holes in X-ray binaries and supermassive black holes in
centers of galaxies may have high values of dimensionless angular
momentum (see e.g., Zhang, Cui \& Chen 1997; Liu, Zhang \& Zhang
2007). Highly spinning black holes are most likely spun up by gas
accretion processes. However, for intermediate mass black holes, it
is currently not clear at all if they should be spinning or not. On
the other hand, black holes with masses around $10^4$ solar masses
are not likely found in accreting binaries. Therefore, if they are
also not located in centers of galaxies where gas densities my be
high enough for fueling significant accretion, they should not be
highly spinning. Here, as a conservative approach we take
$A=0.1$, which leads to $A^2f(A)=0.02$.

Eq.~(12) shows that $E_{\rm pulse}$ is a function of
$\hat{r}_{\rm in}^{-57/64}$. Therefore, the maximum and minimum value
for $E_{\rm pulse}$ will be acquired when the seed fields anchored in the
disk at $\hat{r}_{\rm in}=1$ and $\hat{r}_{\rm in}=\hat{r}_{\rm sp}$,
respectively. Immediately, we derive the maximum value of
$E_{\rm pulse}$ by setting $\hat{r}_{\rm in}=1$ in Eq. (12),
\begin{eqnarray}
E_{\rm pulse,max}= 5.38\times 10^{48}\alpha^{-1}M_5\, {\rm ergs}.
\end{eqnarray}
The minimum value of $E_{\rm pulse}$ can be derived by setting
$\hat{r}_{\rm in}=2.52\dot{m}$ in Eq. (12),
\begin{eqnarray}
E_{\rm pulse,min}= 2.36\times 10^{48}\alpha^{-1}M_5\dot{m}^{-57/64}\, {\rm
ergs} .
\end{eqnarray}
The average value of $E_{\rm pulse}$ for each pulse can be estimated as,
\begin{eqnarray}
E_{\rm pulse,ave}&=&\frac{\int_1^{\hat{r}_{\rm sp}}
E_{\rm pulse}d\hat{r}_{\rm in}}{\int_1^{\hat{r}_{\rm sp}}d\hat{r}_{\rm in}}\nonumber\\
 &=& 4.92\times
10^{49}\alpha^{-1}M_5\frac{(2.52\dot{m})^{7/64}-1}{2.52\dot{m}-1}\,\, {\rm ergs} .
\end{eqnarray}

\subsection{Application to GRB 060614}

We suggest that GRB 060614 could be produced by the tidal disruption
of a solar type star ($m_*=1, r_*=1$) by an IMBH. Here we give a
rough estimate for the parameters involved in this case. As
described in Section 2, the first short episode of emission lasting
for $\sim 4$ s is actually composed of $\sim 5$ mini-pulses. We
notice that the durations of these mini-pulses can be as short as
$\sim 0.6$ s. According to our Eq. (10), we estimate the mass of the
IMBH as $M_5 \approx 0.2$. Taking a typical value of $\alpha = 0.1$,
we then calculate the theoretical duration of the GRB as $t_{\rm
duration} \approx 100$ s from  Eq. (11). This is in good agreement
with the observed duration of $\sim 102$ s.

Assuming that the GRB happens when the IMBH is roughly accreting at
the Eddington rate of $\dot{m}=1$, we can further calculate the
average pulse energy (see Eq. (15) ) as $E_{\rm pulse,ave} \approx
6.88 \times 10^{48}$ ergs. The exact number of mini-pulses is not
easy to determine from observations (it depends too strongly on the
timing resolution of the light curve.). From our Eq. (9), we assume
that there are $N_{\rm tot} \sim 30$ mini-pulses, then we get the
total burst energy as $E_{\rm tot} = N_{\rm tot} E_{\rm pulse,ave}
\approx 2.06 \times 10^{50}$ ergs. Assuming a beaming factor of
$\sim 100$ (Gehrels et al. 2006), it will correspond to an isotropic
kinetic energy of $\sim 2 \times 10^{52}$ ergs, which is consistent
with the observed energetics of GRB 060614 (see Section 2).

As noted in Section 2, there is a hint of a $9\,s$ periodicity
between 7 and 50 s in the $\gamma$-ray light curve \citep{Geh06}.
Such a quasi-periodic oscillation can be naturally explained in our
model. We suggest that this periodicity should be connected with the
Kepler motion at $R_{\rm ms}$. In fact, the Kepler period is
\citep{Sha73}
\begin{equation}
t_{\rm K} = 2 \pi \sqrt{\frac{R_{\rm ms}^3}{G M_{\rm bh}}} \approx
50 \hat{r}_{\rm ms}^{3/2}M_5 \, {\rm s} .
\end{equation}
Taking $M_5=0.2$, we find that $t_{\rm K} \sim 10\,s$, in good
accordance with the observed 9-s periodicity. It thus seems clear
that the falling of the blobs into the black hole is neither uniform
nor completely unsystematic. They seem to fall in group and the
falling is modulated by the Keplerian motion. However, the detailed
modulation mechanism is still largely uncertain and needs further
investigations.

Our model can also naturally explain other basic features of GRB
060614. For example, it needs not to be associated with a supernova
and need not to reside in an active star forming galaxy. It also
needs not to be at the center of its host galaxy, since the black
hole involved is not a super-massive one. In our model, the highly
collimated outflow can also be naturally launched via the BZ
process.

\section{Event rate}

It is interesting to note that another marginally long burst, GRB
060505, with a duration of $\sim 4$ s, was also not accompanied by
supernova emission (Fynbo et al. 2006). Thus we currently have two
confirmed GRBs that are of our interest. Based on this fact, we can
give a rough estimate for the observed event rate of such a
phenomenon. Till the end of March 2008, about 600 GRBs have been
well localized (Greiner 2008). Optical afterglows have been detected
from 233 events of them, with redshifts measured in 141 cases.
According to an earlier study by Zeh, Klose, and Hartmann (2004),
supernova signature is most likely detectable for a GRB with
redshift $z < 0.7$, if it is really associated with a supernova. For
GRBs with redshifts $z > 0.7$, the signature may be too weak to
detect. Taking $z < 0.7$ as a criterion, we then have a sample of 32
GRBs, among them are GRBs 060505 ($z = 0.089$) and 060614 ($z =
0.125$). It hints that about 6\% GRBs are GRB 060614-like. Note that
the local GRB-XRF (X-ray flash) birth rate is $\sim$ (0.5 ---
2)Gpc$^{-1}$ yr$^{-1}$, or $\sim$ (0.025 --- 0.1) Myr$^{-1}$
galaxy$^{-1}$ (Schmidt 2001; Zhang \& M\'esz\'aros 2004). We then
estimate the birth rate of GRB 060614-like event as $\sim$ (0.03 ---
0.12)Gpc$^{-1}$ yr$^{-1}$, or $\sim$ (0.0015 --- 0.006) Myr$^{-1}$
galaxy$^{-1}$. The above calculation is based on the assumption that
the GRB emission is isotropic. If we assume a true-to-observed
beaming correction factor of 100 --- 1000, then the local birth rate
of GRB 060614-like event will be amplified to $\sim$ (3 ---
120)Gpc$^{-1}$ yr$^{-1}$, or $\sim$ (0.15 --- 6) Myr$^{-1}$
galaxy$^{-1}$.

In our model, the birth rate of IMBH-induced GRBs in a typical
galaxy can be calculated as
\begin{equation}
\Gamma = \kappa \Gamma_{\rm t} N_{\rm IMBHs},
\end{equation}
where $N_{\rm IMBHs}$ is the total number of IMBHs in the galaxy,
$\kappa$ is the number ratio of IMBHs whose masses are suitable for
producing a GRB, and $\Gamma_{\rm t}$ is the mean tidal disruption
rate of a solar type star by an IMBH.  We believe that to produce a GRB,
the mass of the IMBH should be larger than $\sim 1 \times 10^4 M_\odot$.
So, $\kappa$ should be the fraction of IMBHs with mass of $\sim
1\times 10^{4}$ --- $10^5 M_{\odot}$ among all the IMBHs with mass of $\sim$
20 --- $10^5M_\odot$. However, direct evidence for the existence
of IMBHs is still lacking, so the three quantities
($\kappa, \Gamma_{\rm t}, N_{\rm IMBHs}$) involved in Equation (17)
are all largely uncertain. Here we can only give some preliminary
discussion.

Although the true existence of IMBHs and their number in a typical
galaxy is uncertain \citep{Map06}, \cite{Vol03} has discussed the
density of IMBHs under the assumption that the IMBHs are born in 3
--- 3.5 $\sigma$ fluctuations collapsing at a given redshift.
Assuming that IMBHs are modeled as a halo population distributed
following a Navarro Frenk \& White (NFW) or a more concentrated
Diemand, Madau \& Moore (DMM) density profile, \cite{Map06} derived
an upper limit for the density of IMBHs via dedicated $N$-body
simulation. Based on their work, the number of IMBHs in the Milky
Way should be $N_{\rm IMBHs} \leq 10^3-10^4$.

The parameter $\kappa$ should seriously depend on the mass function
of IMBHs. However, since the formation mechanism of IMBHs and their
evolution after birth are completely uncertain, we actually cannot
give any robust estimate for $\kappa$. Also, the environment that
IMBHs reside in is unknown, so that a reliable estimate for the
tidal disruption rate ($\Gamma_{\rm t}$) is again impossible.

Although it is still impossible to estimate the rate of producing
GRB 060614-like GRBs in our model, we believe that this model
provides a viable GRB production mechanism. In the future, when more
and more GRB 060614-like events are observed, more stringent
constraints on $\kappa$, $\Gamma_{\rm t}$ and $N_{\rm IMBHs}$ may be
available, and a valuable new window will be open for the study of
IMBHs.

\section{Discussion and conclusion}

We have shown that the tidal disruption of a solar type star by an
IMBH with a mass of $2.0\times 10^{4}M_\odot$ can be a possible
mechanism for the special event of GRB 060614, which is a nearby
long burst but is not associated with a supernova. We argue that the
powerful energy extracted via the BZ process is enough to trigger a
GRB, when the black hole is accreting at the Eddington rate.
The basic observed features of GRB 060614 can all be reasonably
explained in our frame work.

The black hole mass considered here is higher for ULXs \citep{Bau06}
and lower for X-ray flares \citep{kom04} via tidal capture and
disruption events. Therefore, no ULXs or X-ray flares could be
observed to be associated with the prompt $\gamma$-ray emission
in this case.

A GRB produced by the tidal disruption of a star by an IMBH may
also leave some hints in its afterglow.
It is interesting to note that an oscillatory phenomenon has been
observed in the optical afterglow of GRB 050922C, with a period of
$\sim$ 7.2 min (Zhilyaev et al. 2007). Zhilyaev et al. (2007)
suggested that GRB 050922C may be resulted from tidal disruption of a
white dwarf star by an IMBH. The observed periodicity in the afterglow
is interpreted as due to the precession of an accretion disc.

\acknowledgements
We thank the anonymous referee for many useful suggestions that
lead to an overall improvement of our manuscript.
We also would like to thank K.S. Cheng, B. Zhang, Y. Z. Fan, Z.
Zheng, Z. G. Dai and X. D. Li for their helpful comments and discussions. This research
was supported by the National Natural Science Foundation of China (Grants 10273011,
10573021, 10433010, 10625313, 10521001, 10733010, 10725313 and 10221001), and by
Chinese Academy of Science through project No. KJCX2-YW-T03.

{}

\end{document}